\documentclass[aps,twocolumn,floats,showpacs]{revtex4}
\usepackage{amsmath,amssymb,graphicx,epsf}

\begin{document}

\title{Antiphase Stripe Order as the Origin of Electron Pockets Observed in 1/8-Hole-Doped Cuprates} 

\author{Andrew J. Millis}
\affiliation{Department of Physics, Columbia University, 538 W 120th St., New York, NY 10027}

\author{M. R. Norman}
\affiliation{Materials Science Division, Argonne National Laboratory, Argonne IL 60439 }

\begin{abstract}
Recent quantum oscillation measurements on underdoped cuprates are shown to be consistent with
the predictions of a mean field theory of the $1/8$ magnetic antiphase stripe order proposed to occur
in high-$T_c$ cuprates. In particular, for intermediate values of the stripe order parameter, 
the magneto-transport is found to be dominated by an {\em electron} pocket.
\end{abstract}
\date{\today}
 \pacs{74.25.Jb, 72.15.Gd, 75.30.Fv}

 \maketitle

\begin{figure}
\centerline{\includegraphics[width=3.0in]{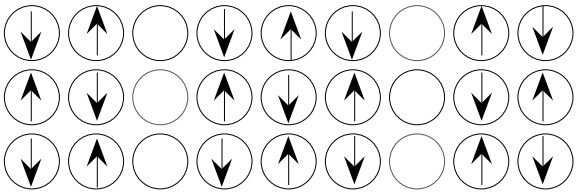}}
\caption{Sketch of the magnetic and charge density in
the  $1/8$ stripe state proposed
for cuprate superconductors. Circles represent copper ions with their surrounding oxygens,
arrows denote the
copper spins, and empty circles indicate the ``charge stripe" (antiphase
domain wall).}
 \label{fig1}
\end{figure}

The interplay between magnetism and superconductivity is an enduring theme in the 
physics of high-temperature copper-oxide superconductivity.  These superconductors are created
by doping antiferromagnetic insulating ``parent compounds", and a wide variety of magnetic phenomena
have been observed in both doped and undoped cuprates.  One of the most interesting magnetic states proposed
for the hole-doped cuprates is the $1/8$ antiphase stripe state. This state has the spin and charge pattern
sketched in Fig.~\ref{fig1}: a modulation of the charge density which has the translation symmetry of the lattice
along one direction (${\hat y}$) and a four unit cell repeat distance along the orthogonal direction
(${\hat x}$). The basic spin pattern is a two-sublattice antiferromagnet, but the lines of minimum charge density are 
taken to be antiphase domain boundaries for the magnetization.  This state was first predicted in Hartree-Fock calculations
of Zaanen and Gunnarsson \cite{Zaanen89}, Machida \cite{Machida}, and  Littlewood and Inui \cite{Inui91}.  It was later 
discussed extensively by Kivelson and collaborators
(for a review, see Ref.~\onlinecite{Kivelson03}).

A non-superconducting stripe phase
has been  unambiguously established as the ground state of
La$_{1.875}$Ba$_{0.125}$CuO$_{4-\delta}$ \cite{Tranquada95} in which the hole density per unit
cell is $1/8$.  A magnetically ordered state (believed to be a stripe phase) can be induced by applying a magnetic
field to the superconducting state of La$_{2-x}$Sr$_x$CuO$_{4-\delta}$ (LSCO) \cite{Lake02} for a range of $x$ around $1/8$.  
Various anomalies associated with  $1/8$ doping have been seen 
clearly in other La$_2$CuO$_4$-derived materials \cite{Sera,Adachi,Lucarelli03},
suggesting that a stripe phase, although perhaps not the ground
state, is very nearby in free  energy and influences measured properties, for example by competing with the superconducting
state.  However, the relevance of stripe phases to
other members of the high-$T_c$ family, in particular the YBa$_2$Cu$_3$O$_{7-\delta}$ (YBCO) 
family of superconductors, has been less clear. 
Magnetic fluctuations are observed in underdoped YBCO, but in the best samples, zero field magnetic order is 
apparently observed only in very underdoped
compounds in the region where superconductivity has already vanished \cite{Yamani07}.  
Transport \cite{Ando99,Ando04}  and magneto-optical
\cite{Rigal04} data have been argued to be indications of density wave or stripe order in moderately-underdoped
YBCO materials, and the ``$60K$ $T_c$ plateau" has also been suggested \cite{Segawa01} to arise from
$1/8$ effects similar to those seen in LSCO.  While these experiments have been suggestive, 
definitive proof of magnetic ordering has been lacking.

Recently, a  very remarkable series of magneto-transport experiments have transformed the situation. In
ultra-pure samples of ortho-II YBa$_2$Cu$_3$O$_{6.5}$ \cite{Doiron07} and YBa$_2$Cu$_4$O$_8$ \cite{Yelland07,Bangura07},
quantum oscillations have been observed at fields above about $50T$. The measured oscillation frequencies suggest
that the signal arises from small Fermi surface ``pockets" such as might be produced by density wave ordering
\cite{Doiron07,Yelland07,Bangura07,Chen07}.
Very recently,  LeBoeuf {\it et al.}~\cite{Tailleferprivate} have 
found  that the   temperature below which the high-field Hall coefficient becomes negative peaks
at a doping of 1/8, providing strong evidence that the quantum oscillations
arise from a Fermi surface reconstruction caused by ``1/8" type ordering.
However,  the sign of the 
measured Hall conductance suggests that the transport arises from {\em electron}
pockets \cite{Tailleferprivate}, whereas
the prevailing consensus is that hole pockets are expected: in particular,
a simple two-sublattice ordering pattern does not yield robust electron pockets \cite{Chen07}
while a large-amplitude stripe ordering would be expected to lead to effectively one dimensional transport characterized by
open Fermi surfaces which would not give rise to magneto-oscillations.  

In this communication we present a theoretical investigation of the Fermi surface and oscillation 
frequency implied by a stripe ordering pattern such as that  sketched in Fig.~\ref{fig1}.  Generically,
complicated Fermi surfaces occur involving open orbits, hole pockets and electron pockets.
However we find that in some intermediate parameter regimes,
the periodic potential associated with the stripe state  can 
produce a Fermi surface consisting of  a set of open (quasi one dimensional) bands, 
along with a single, simple electron pocket. The open orbits will make no
contribution to the magneto-oscillations, while the electron pocket exhibits properties 
which are are  in qualitative agreement 
with the magneto-oscillation data.   We therefore argue that the recent magneto-transport data support the idea
that stripe phases (competing with superconductivity, and most probably
field-induced) are generic to hole-doped cuprates. 

We consider the canonical tight-binding model of electrons hopping on a square lattice
which is believed to describe the Fermi surface of the cuprates:
\begin{eqnarray}
\varepsilon_k&=&-2t\left(\cos k_x+\cos k_y\right) +4t^{\prime}\cos k_x\cos k_y
\nonumber \\
&&-2t^{\prime\prime}\left(\cos2k_x+\cos2k_y\right) -\mu
\label{dispersion} 
\end{eqnarray}
where following Ref.~\onlinecite{Andersen95} we choose 
$t=0.38eV$, $t^{\prime}=0.32t=0.122eV$ and $t^{\prime\prime}=0.5t^{\prime}=0.061eV$, 
noting that $\mu$ is the chemical potential. We express momenta in $\pi/a$ units, with $a$ the lattice
parameter of the underlying square lattice.

We assume that electrons moving in this band structure are scattered by potentials with the periodicity
of the stripe order. As can be seen from Fig.~\ref{fig1}, the stripe  state has a unit cell containing $8$ sites
of the underlying lattice.  Fourier transforming the spins, we obtain a potential $V_n$ 
connecting the state ${\bf k}$ with those at ${\bf k}\pm n{\bf Q}$  with   ${\bf Q}=(3\pi/4,\pi)$. 
We expect that the term $V_1$ will
be the dominant spin-derived scattering term,  with a weaker third harmonic $V_3$ at 3${\bf Q}_\pm$.
There will be a charge component $V_2$ at 2${\bf Q}_\pm$ with a weaker second harmonic
at 4${\bf Q}_\pm$.  We have found that the influence of $V_3$ and $V_4$ on our results is minor;
these terms will be neglected here.
Relabeling the spin potential $V_1$ as $V$ and the charge potential $V_2$ as $V_c$,
%The primitive translation vector may be taken to be ${\bf b}=(4 {\hat x},{\hat y})$ with
%${\hat x},{\hat y}$ the lattice vectors of the underlaying square lattice.  The reduced translation symmetry
%implies a potential  $V$ which connects an electronic state
%at wavevector ${\bf k}$ to states at wavevectors ${\bf k}\pm {\bf Q}_\pm={\bf k}\pm(\pi/4,\pi)$. 
%If we neglect the spin structure, the state is characterized by the shorter translation vector
%${\bf c}=(4{\hat x},0)$. We therefore expect in addition a potential $V_c$
%arising from the charge modulation with wavevector  ${\bf K}=(\pi/2,0)$.
%We assume that the spin scattering is the dominant effect, so that $V>V_c$; for 
%most of our calculations we take $V_c=V/4$, but the precise
%ratio is not important. 
the resulting Hamiltonian may be written as an $8\times 8$ matrix for ${\bf k}$ in the first Brillouin zone of the ordered state
\begin{widetext}
\begin{eqnarray}
H=
\left(\begin{array}{cccccccc}
\varepsilon_k & V_c & 0 & V_c & 0 & V & V & 0 
\\V_c& \varepsilon_{k+(\frac{1}{2},0)} & V_c & 0 & 0 & 0 & V & V 
\\0 & V_c & \varepsilon_{k+(1,0)}  & V_c & V & 0 & 0 & V 
\\V_c & 0 & V_c & \varepsilon_{k+(\frac{3}{2},0)} & V & V & 0 & 0 
\\0 & 0 & V & V & \varepsilon_{k+(\frac{1}{4},1)} & V_c & 0 & V_c 
\\V & 0 & 0 & V & V_c & \varepsilon_{k+(\frac{3}{4},1)} & V_c & 0 
\\V & V & 0 & 0 & 0 & V_c& \varepsilon_{k+(\frac{5}{4},1)}   & V_c
\\0 & V & V & 0 & V_c & 0 & V_c & \varepsilon_{k+(\frac{7}{4},1)}\end{array}\right)
\label{H}
\end{eqnarray}
\end{widetext}
%(note that $V_4$ would replace $0$ in the diagonal blocks, and $V_3$ would replace $0$ in the off-diagonal ones, if these potentials were included).
We have diagonalized this matrix and determined the Fermi surface for various dopings
and $V_n$ (note that the chemical potential has to be adjusted for each set of $V_n$ to preserve the doping).
We find that the Fermi surfaces  are most easily visualized if the results are plotted not in the reduced Brillouin zone,
but in a quadrant of the full square lattice zone $(0<k_x<1),(0<k_y<1)$.

\begin{figure}
\centerline{\includegraphics[width=3.4in]{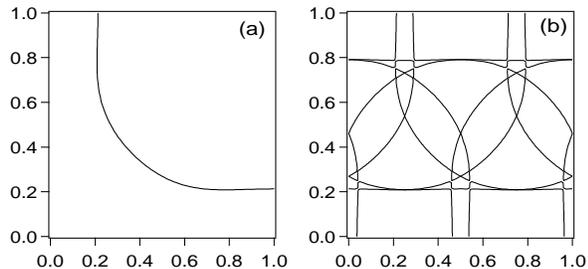}}
\caption{(a) Fermi surface from Eq.~1, corresponding to a hole density of 1/8,
plotted in the first quadrant of the Brillouin zone of the underlying square lattice
(note that $\pi/a$ momentum space units are used in all figures).
(b) Fermi surface in (a) plus its images under translation by 
the stripe order (multiples of ${\bf Q_\pm}$), equivalent to Eq.~\ref{H} with $V=V_c=0$.}
\label{fig2}
\end{figure}

To understand the results, it is useful to examine first the non-ordered ($V=V_c=0$) case.
Fig.~\ref{fig2}a shows the Fermi surface from Eq.~\ref{dispersion} at 
$1/8$ hole doping.  The Fermi surface is a large hole surface centered
at the point $(1,1)$.
Fig.~\ref{fig2}b shows this Fermi surface, along with the surface translated
by multiples of ${\bf Q_\pm}$. The result is a highly complicated set of bands. 
The key feature, however,
is the nearly flat segments. These arise from the portions of the Fermi surface 
which approach the zone boundaries 
$k_x=1/k_y=1$.  We shall see that the scattering potential implied by the stripe order
reconnects these into an electron pocket, which exists over a wide range of parameter values.

\begin{figure}
\centerline{\includegraphics[width=3.4in]{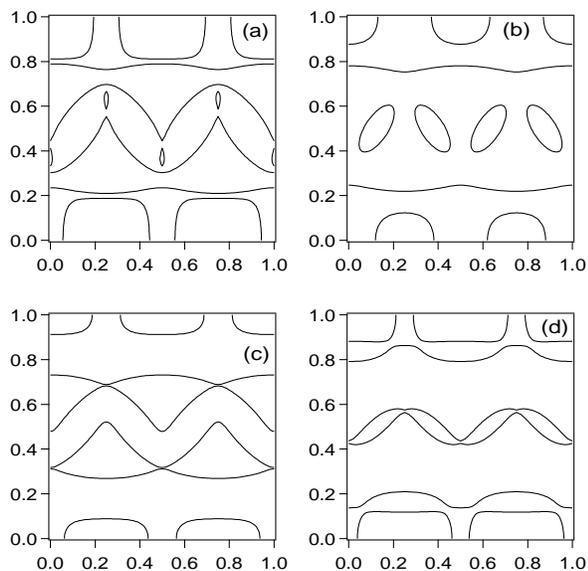}}
\caption{Fermi surfaces from Eq.~\ref{H} with
a hole doping of 1/8 plotted in the first quadrant of the Brillouin zone 
of the underlying square lattice: (a) $V=0.1eV$, $V_c=0$, (b) $V=0.2eV$,
$V_c=0$, (c) $V=0.2eV$, $V_c=0.15eV$, and (d) $V=0.2eV$, $V_c=-0.2eV$.}
\label{fig3}
\end{figure}

The four panels of Fig.~\ref{fig3} shows the effect of the stripe order. Panel $a$ shows the
effect of a weak spin potential with no charge potential.  The resulting Fermi surface is reconstructed
and begins to reveal the one dimensional structure
associated with motion along the stripe direction. The electron pocket has also emerged,
although its size is considerably larger than that observed in Ref.~\onlinecite{Doiron07}.

Panel $b$  shows the Fermi surfaces implied by the stronger coupling
$V=0.2eV$ (with $V_c=0$). The increase in $V$ shrinks the electron pocket and separates it from
the one dimensional band.  In addition, the ``zig-zag" band has pinched off, leading to
hole pockets.  Reducing the hole doping to 1/10 or moderately increasing $V$ causes these hole
pockets to decrease
and ultimately disappear, leaving only electron pockets. 
For yet larger $V$, the electron pockets vanish and only the quasi one dimensional bands remain.

Panels $c$ and $d$ explore the effect of adding a potential due to the charge order. Fourier transformation of 
the standard stripe ordering pattern implies $V_c>0$ as may be understood from the lower {\em electron} density on
the charge stripe row, but for completeness we show both signs.  In both cases the effect of $V_c$ is to reconnect
the hole pockets into one dimensional bands; the $V_c<0$ case tends to eliminate
them more rapidly than does $V_c>0$.

\begin{figure}
\centerline{\includegraphics[width=3.4in]{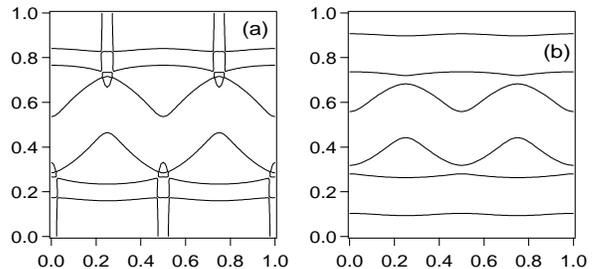}}
\caption{Fermi surface from Eq.~\ref{H} with $V=0$ and (a) $V_c=0.1eV$ and (b) $V_c=0.2eV$,
for $1/8$ hole doping, plotted in the first quadrant of the Brillouin zone of the underlying square lattice.}
\label{fig4}
\end{figure}

One question which arises is the relative effects of magnetic and charge order. To investigate this
point, we have re-computed the Fermi surface for the case $V=0$ with varying $V_c$ (i.e., assuming
that the magnetic order has a negligible effect on the electron dynamics). A representative Fermi surface 
is shown in Fig.~\ref{fig4}a.  We see that in the absence of spin order, the pockets
have a very extreme aspect ratio. As the potential
is made stronger, the aspect ratio of the pocket increases until it collapses into a single line and
vanishes (Fig.~\ref{fig4}b).  And in this purely charge case, the pocket is a hole
pocket. 
%On the other hand, calculations in which $V_c$ is set to zero but $V$ is retained
%lead to a Fermi surface not materially different from those shown for the $V_c=V/4$ case.
We conclude that the potential produced by spin ordering plays the essential role
in the formation of the observed electron pocket.  
%This would be consistent with the field induced nature of the 1/8 state in the YBCO case.
%I AM NOT SURE I BUY THIS: THE GENERAL IDEA IS THAT YOU GET FIELD-INDUCED ORDER BECAUSE SC IS SUPPRESSED. IF ANYTHING FIELD WOULD BE WORSE FOR MAGNETIC THAN FOR CHARGE ORDER

We next examine  in more detail the  pockets implied by our calculations.
Within the model,  the size of the electron pockets depends strongly on $V$; the value of $V$ may therefore
be adjusted to produce any area desired at a given doping.
%The parameters of
%Fig.~\ref{fig3}d have been adjusted to give an electron pocket of size
%We choose the value  $V=0.4eV$ as an example of a pocket which is neither
%so large that it crosses the quasi one dimensional bands, nor so small that it almost does not exist. The area of the pocket
Since there are 8 magnetic zones per square lattice zone, and there is 1 electron
pocket per magnetic zone,
a pocket with size
$7.6\%$ of the area of the first quadrant of the square lattice Brillouin zone, as
seen in Ref.~\onlinecite{Doiron07},
corresponds to an occupation factor
of 0.038 electrons per copper atom (noting that the magnetic state retains a Kramers degeneracy).
%roughly consistent with the quantum oscillation value
%of $7.6\%$ \cite{Doiron07}.

The  electron pocket, centered at the magnetic zone boundary, $(1/4,0)$, has a cyclotron mass
which varies from 0.5 to 1.0 for the various cases shown in Fig.~\ref{fig3}.
The hole pocket in Fig.~\ref{fig3}b has a lighter cyclotron mass of 0.3.  Therefore, if such a hole pocket
existed, it would be very apparent in any magneto-oscillation measurement.
%whereas the hole pockets are elliptica with a 
%mass ratio of about $4-1$ and a cyclotron frequency (geometric mean of the two masses) corresponding to the much smaller mass
%$m=0.12m_e$. The mass depends relatively weakly on the value of $V$ in this regime:
%as $V$ is increased, the main effect is to raise the bottom of the electron band until it passes through the
%Fermi energy These light masses suggest that the hole pockets in this calculation would make an important contribution to
%the dc transport, although they might not be seen in quantum oscillation measurements.
%The masses do depend on the value of the charge scatterng amplitude $V_c$. For the electron pockets shown in Fig. 3c
%we find again a mass ratio of about $4-1$ and a cyclotron mass of about $1.2m_e$. 
%\begin{equation}
%E^*=\frac{p_x^2}{2m_x}+\frac{p_y^2}{2m_y}-E_F^*
%\label{epocket}
%\end{equation}
%where $p_{x,y}$ are momenta measured relative to $(1/4,0)$. 
%Restoring units, we find $E_F^*=0.063eV$, $m_x=0.5m_e$ and $m_y=1.4m_e$,
  %The mass measured in a   quantum oscillation measurement is $\sqrt{m_xm_y}=0.8m_e$.
The measured cyclotron mass is $1.9m_e$ \cite{Doiron07}.  We suggest that the difference between
the calculated and measured masses is due in large part to the renormalization of  
band theory inferred from other measurements.
For example, the nodal velocity measured in
angle-resolved photoemission experiments is about $1.6eV \AA$  \cite{Damascelli03}, which is 
reduced from the calculated  band velocity of $3.8eV \AA$ from Eq.~1 by a factor of $2.4$.
Such a renormalization would imply a bare cyclotron mass of 0.8, which is within the range of
the values we find for the electron pocket.
%Applying this renormalization brings the mass into remarkable consistency
%with experiment.

In summary, we have shown that at intermediate coupling, a periodic potential such as would be produced
by a magnetic antiphase domain ``stripe" structure produces electron pockets which are consistent with experiment, 
if the energy scales are renormalized from band theory by a factor similar to that
observed from angle-resolved photoemission
experiments. The results are sensitive to the details of the scattering potential: in particular, for stronger potentials $V \gtrsim 0.3eV$,
only open orbits associated with one dimensional transport along the stripe direction exist, whereas
for a range of combinations of spin and charge potentials, both electron and hole pockets are found,
though the latter exist for a much narrower range of parameters than the former. 
We emphasize, though, the generic feature of our results, namely that the spin potential associated
with antiphase stripe order produces robust  electron pockets in hole-doped high-$T_c$ superconductors.
These results  confirm the interpretation of Ref.~\cite{Doiron07}
that the quantum oscillations
imply the existence of a density wave state at high magnetic fields in the YBCO cuprates, and indeed
confirm the identification of the state as the $1/8$ magnetic antiphase stripe state discussed by many workers.
However, we note that our results (and, more fundamentally, the identification of electron pockets) 
imply that
the ordering is not strong enough to confine the electron motion completely to one dimensional paths.
Our findings suggest several further directions for research.  
X-ray or elastic 
neutron scattering experiments should be performed at high fields 
to confirm this identification. Further, as noted by Ando {\it et al.}~\cite{Ando04}, stripes in the YBCO materials
are expected to be aligned along the chain direction; thus in single-domain samples, the
stripe domains may all be aligned, leading to an increase in the anisotropy. Optical and 
magneto-optical experiments may reveal evidence of the excitation gap implied by the stripe order.
A future paper will present a more comprehensive theoretical investigation,
including the effects of bilayer coupling and oxygen ordering,
as well as computing in detail the magneto-transport and magneto-optical properties of the various states. 

We thank Louis Taillefer for sharing unpublished data \cite{Tailleferprivate} with us,
and he and Steve Kivelson for helpful discussions.
This work was performed at the Aspen Center for Physics.  AJM was supported by 
NSF-DMR-0705847 and MRN by the U.S. DOE, Office of Science,
under contract DE-AC02-06CH11357.

\end{document}